\documentclass[prl,twocolumn,twoside,showpacs,superscriptaddress]{revtex4}
\usepackage{graphics,amssymb,amsmath,epsf}
\epsfclipon

\usepackage{xspace}

\begin{document}

\title{Infrared propagators of Yang-Mills theory from perturbation theory}

\author{Matthieu Tissier}
\affiliation{Laboratoire de Physique Th\'eorique de la Mati\`ere Condens\'ee,
Universit\'e Pierre et Marie Curie, 4 Place Jussieu 75252 Paris CEDEX 05, France}
\author{Nicol\'as Wschebor} 
\affiliation{Instituto de F\'{\i}sica, Facultad de Ingenier\'{\i}a, Universidad de la Rep\'ublica, 
J.H.y Reissig 565, 11000 Montevideo, Uruguay}

\begin{abstract}
  We show that the correlation functions of ghosts and gluons for the
  pure Yang-Mills theory in Landau gauge can be accurately reproduced
  for all momenta by a one-loop calculation. The key point is to use a
  massive extension of the Faddeev-Popov action. The agreement with
  lattice simulation is excellent in $d=4$. The one-loop calculation
  also reproduces all the characteristic features of the lattice
  simulations in $d=3$ and naturally explains the pecularities of the
  propagators in $d=2$.
\end{abstract}
\pacs{ 12.38.-t, 12.38.Aw, 12.38.Bx,11.10.Kk}
\maketitle The infrared (IR) physics of strong interaction is well
described today by lattice simulations of Quantum Chromodynamics
(QCD).  This tool is now commonly used to determine the spectrum of
particles, cross sections and other physical observables (see for
example~\cite{Montvay94}). The analytical (or semi-analytical)
approaches have not reached such a high level of development, mainly
because the standard perturbative approach breaks down at low
energies. This calls for more sophisticated techniques, and a good
guiding principle in developing them is to compare their predictions
with lattice simulations. Unfortunately, the simplest quantities that
can be computed in analytical approaches are non-gauge-invariant and
therefore require to fix the gauge.  For this reason, a considerable
amount of work has been performed in the last few years to study
gauge-fixed versions of QCD in the lattice.

The simplest quantities that can be analytically studied are the
2-point correlation functions in Landau gauge, and most of the
gauge-fixed simulations have focused on these quantities.  In the case
of pure gluodynamics (with no quarks), some facts are now clearly
established. First, the gluon propagator does not diverge in the IR
but tends to a positive constant for
$d>2$~\cite{Cucchieri,Bogolubsky09,Dudal10} and to zero in
$d=2$~\cite{Cucchieri_08,Maas_07}. The ghost propagator is divergent
in the IR limit with an enhancement when compared to the free
propagator: the dressing function (i.e. the propagator times momentum
squared) is monotonically decreasing with momentum. It seems to
approach a finite positive constant in the IR for $d>2$ and diverges
in this limit for $d=2$. It is also well documented that the
K\"{a}ll\'en-Lhemann spectral function associated with the gluon
propagator is not definite positive~\cite{Bowman07,Cucchieri_04}.

Let us now recall the various analytical approaches that have been
used to determine these correlation functions.  The standard
perturbation theory in the framework of Faddeev-Popov (FP)
gauge-fixing, as is well-known, is unable to access the IR limit of
the theory because it presents a Landau pole.  This may be related to
the fact that the FP procedure does not fix completely the gauge
because of the Gribov ambiguity: there exists in general several gauge
transformed configurations (Gribov copies) that satisfy a gauge
condition~\cite{Gribov77}. A line of investigation has been developed
to restrict the functional integral in order to take into account only
a subset of the Gribov copies (hopefully, only one). This leads to the
Gribov-Zwanziger model~\cite{Gribov77,Zwanziger89,Zwanziger92} and
some variants of it~\cite{Dudal08}. The IR propagators have also been
studied by using Schwinger-Dyson and Non-Perturbative Renormalization
Group equations. In these approaches, one solves a truncated version
of an infinite set of coupled equations for the vertex
functions. Depending on how one implements these ideas, two families
of solutions have been found: i) the scaling
solution~\cite{Alkofer00,Fischer04}, where the gluon propagator goes
to zero in all dimensions and the ghost propagator is more singular
than the bare one in the IR, and ii) the decoupling
solution~\cite{Boucaud06} where the propagators have behaviors in
qualitative agreements with the lattice simulations. We note at this
level that all these approaches lead to quite involved calculations,
with in some cases an important numerical part.

In this letter we take a more pragmatic point of view. We do not try
to find a gauge-fixed theory that would be justified from first
principles, but propose a minimal modification of the FP action that
can account for the lattice simulation results. Of course, this
phenomenological approach can only be motivated a-posteriori, if it
describes in a satisfactory way the simulation results. Our main guide
is the observation that the gluon propagator tends to a finite
positive value in the IR for $d>2$. We propose to impose this property
at the tree level by adding a mass term for the gluon in the FP action
\footnote{If we choose not to modify the field content of the theory,
  mass terms are the only local and renormalizable modifications of
  the FP action that do not affect the UV behavior of the theory.}. We
do not change the ghost sector since the ghost propagator is found to
be IR divergent in the simulations. This leads us to consider the
Landau-gauge FP euclidean Lagrangian supplemented with a gluon mass
term:
\begin{equation}
  \label{eq_lagrang}
  \mathcal L=\frac 14 (F_{\mu\nu}^a)^2+\partial _\mu\overline c^a(D_\mu
  c)^a+ih^a\partial_\mu A_\mu^a+\frac {m^2}2 (A_\mu^a)^2 
\end{equation}
where $(D_\mu c)^a=\partial_\mu c^a+g f^{abc}A_\mu^b c^c$ and the
field strength $F_{\mu\nu}^a=\partial_\mu A_\nu-\partial_\nu
A_\mu+gf^{abc}A_\mu^bA_\nu^c$ are expressed in terms of the coupling
constant $g$. The Lagrangian~(\ref{eq_lagrang}) corresponds to a
particular case of the Curci-Ferrari model~\cite{Curci76}. At the tree
level, the gluon propagator is massive and transverse in momentum
space:
\begin{equation}
\label{eq_propag_bare}
  G_{\mu\nu}^{ab}(p)=\delta^{ab} P^\perp_{\mu\nu}(p)\frac 1{p^2+m^2}
\end{equation}
with $P^\perp_{\mu\nu}(p)=\delta_{\mu\nu}-p_\mu p_\nu/p^2$.  It is
interesting to note that the spectral density associated with the
propagator~(\ref{eq_propag_bare}) is positive and therefore there is
no violation of positivity at the tree level. We conclude that
violation of positivity, if it exists in this model, is caused by
fluctuations.

Actually, the gluon propagator observed in the lattice is not
compatible with the bare propagator~(\ref{eq_propag_bare}) and we will
show below that, by including the one-loop corrections, one obtains
propagators for gluons and ghosts that are in impressive agreement
with those obtained in the lattice in $d=4$ (including positivity
violations) and that reproduce at the qualitative level the results
for $d=3$. Let us mention that a mass term has been used to improve
perturbative QCD results in order to reproduce the phenomenology of
Strong Interactions~\cite{Natale09}. Moreover, there are successful
confinement models~\cite{Cornwall79} that use actions including a
gluon mass term.

When analyzing the model described above, we must face the problem
that the mass term breaks the BRST symmetry~\cite{BRST} which is very
important in the perturbative analysis.  This symmetry has the form
\begin{equation}
\label{BRST}
\begin{array}{ll}
  \delta A_{\mu}^a= \eta\, (D_\mu c)^a, &\delta c^a= - \eta\,\frac g 2 f^{abc} c^b c^c, \\
  \delta \bar c^a= \eta\, i h^a, &\delta i h^a = 0,
\end{array}
\end{equation}
where $\eta$ is a global grassmanian parameter.  The BRST symmetry is
in general used to prove the renormalizability of the theory. However,
the breaking of the BRST symmetry by the mass term is soft and
therefore does not spoil renormalizability~\cite{Curci76}. The BRST
symmetry is also used to reduce the state space to the physical space,
in which the theory is unitary (at least at the perturbative level)
and the breaking this symmetry spoils the standard proof of
unitarity. This problem is actually common to essentially all methods
that try to go beyond the standard perturbation theory (as the
Gribov-Zwanziger model) because they all break the standard BRST
symmetry. In this respect, the model considered here is not in a worse
position than other approaches considered in the field. We must stress
that this model is equivalent to the standard FP model in the
ultraviolet limit $p \gg \Lambda_{QCD}$ if $m \sim
\Lambda_{QCD}$. This means that in the domain of validity of standard
perturbation theory, the model is as unitary as QCD. The unitarity of
the model in other momentum regimes is of course an important open
problem, as it is in all gauge fixings in which standard BRST symmetry
is broken.

\begin{figure}
\epsfxsize=7.4cm
\epsfbox{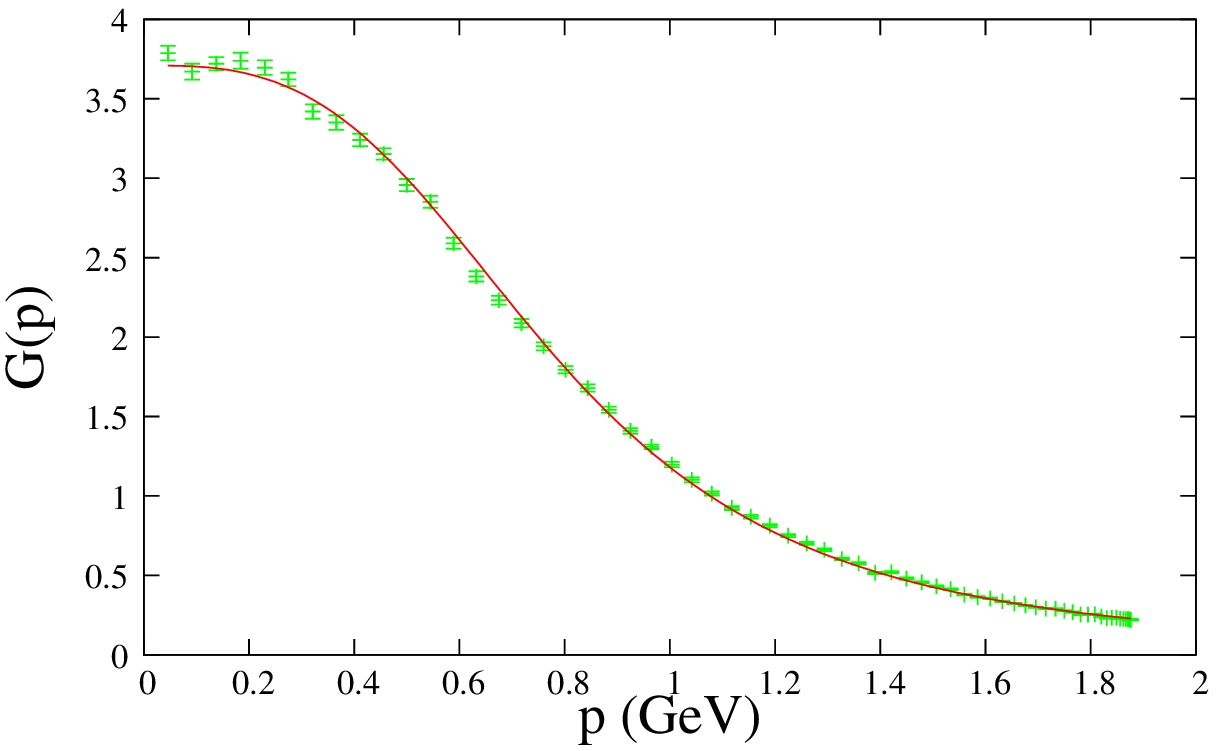}
\epsfxsize=7.4cm
\epsfbox{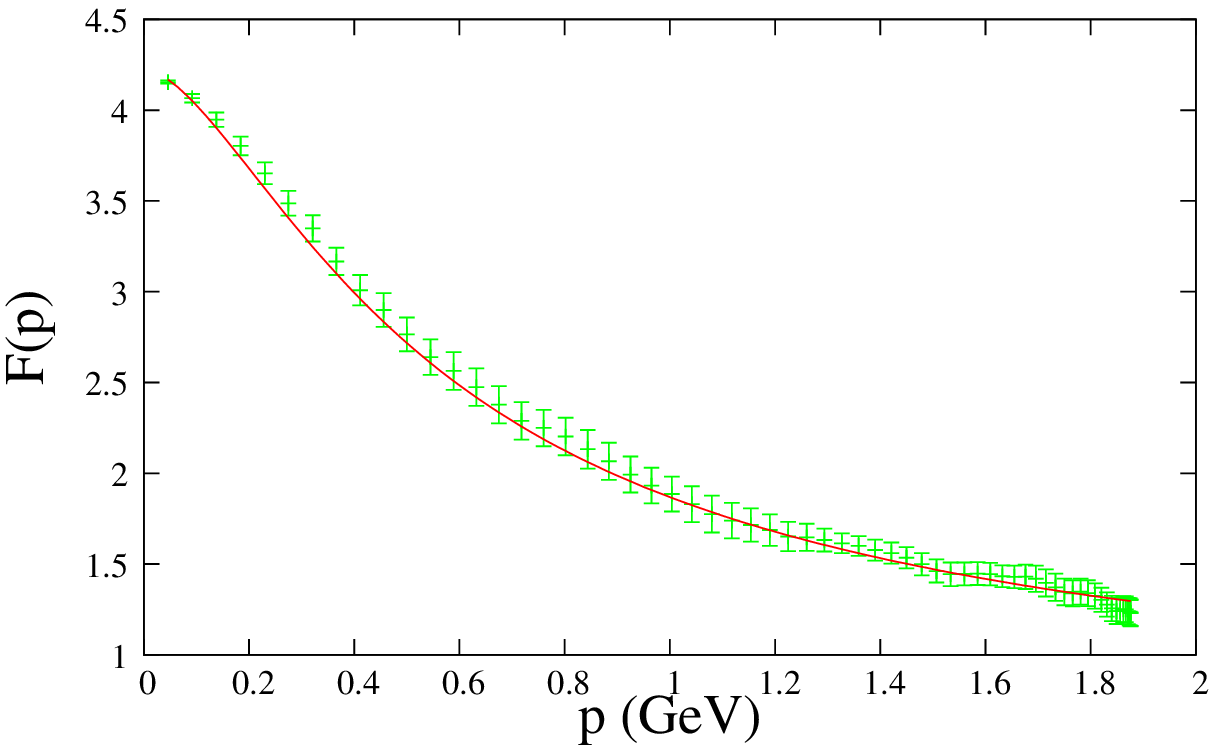}
\caption{\label{fig2}Four-dimensional correlation functions for
  SU(2) gauge group. The results of the present work (red curve) are
  compared with lattice data of~\cite{Cucchieri} (green
  points).  Top figure: gluon propagator. Bottom figure: ghost
  dressing function.}
\end{figure}
\begin{figure}
\epsfxsize=7.4cm
\epsfbox{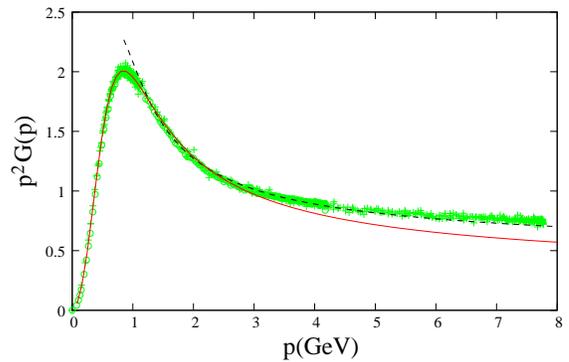}
\caption{\label{fig3}Four-dimensional gluon propagator for SU(3)
  gauge group times $p^2$.  The results of the present work (red curve)
  are compared with lattice data of~\cite{Bogolubsky09} (green open
  circles) and~\cite{Dudal10} (green crosses). The (black) dashed
  curve is the ultraviolet-improved curve obtained by the
  renormalization group.}
\end{figure}
The model with Lagrangian~(\ref{eq_lagrang}), as a particular case of
the Curci-Ferrari model, has a pseudo-BRST symmetry (not nilpotent)
that has the same form as the standard BRST~(\ref{BRST}) except for
the $h$ variation which reads $\delta i h^a = \eta\, m^2 c^a$. On top
of this symmetry, the Lagrangian has all the standard symmetries of
the FP action for the Landau gauge. This includes the shift in
antighost $\bar c\to \bar c +cst.$, a symplectic
group~\cite{Delduc89}, and four gauged supersymmetries recently
found~\cite{Tissier08}. As a consequence, the mass~\cite{Dudal02} and
coupling constant~\cite{Taylor71} renormalization factors (even in
presence of the mass term~\cite{Tissier08}) are fixed in terms of
gluon and ghost field renormalizations.

We present now the 1-loop calculation of the propagators, which
requires the calculation of four Feynman diagrams. It is convenient to
parametrize the gluon $G_{\mu\nu}^{ab}(p)$ and the ghost $G^{ab}(p)$
propagators in the form:
\begin{equation}
 G^{ab}(p)=\delta_{ab}F(p)/p^2, \hspace{.2cm} G_{\mu\nu}^{ab}(p)=P^\perp_{\mu\nu}(p) \delta_{ab} G(p).
\end{equation}
The $F(p)$ is known as the ghost dressing function and the scalar
function $G(p)$ will be refered to as the gluon propagator below.  We
choose the following renormalization conditions:
\begin{align}
\label{rencond}
&G(p=0)=1/m^2, \hspace{.4cm} G(p=\mu)=1/(m^2+\mu^2),\nonumber\\
&F(p=\mu)=1.
\end{align}
We use the gluon-ghost vertex in the Taylor scheme~\cite{Taylor71} for
the coupling constant $g$.

We consider first the 4-dimensional case. The one-loop result for the
renormalized functions $F(p)$ and $G(p)$ (imposing the renormalization
prescriptions~(\ref{rencond})) are:
\begin{align}
\label{4dprops}
&G^{-1}(p)/m^2=s+1+ \frac{g^2 N s}{384 \pi ^2} \Big\{111s^{-1}-2 s^{-2}
\nonumber\\
&\hspace{.3cm}+(2-s^2)\log s+2(s^{-1}+1)^3\left(s^2-10 s+1\right) \log(1+s)\nonumber\\
&\hspace{.3cm}+(4 s^{-1} +1)^{3/2}
   \left(s^2-20 s+12\right)\log \left(\frac{\sqrt{4+s}-\sqrt{s}}{\sqrt{4
   +s}+\sqrt{s}}\right) \nonumber\\
&\hspace{1.3cm}-(s\to \mu^2/m^2)\Big\} \nonumber \\
&F^{-1}(p)=1+\frac{g^2 N}{64 \pi ^2} \Big\{-s \log s +(s+1)^3 s^{-2} \log(s+1)\nonumber\\
&\hspace{1.3cm} - s^{-1}-(s\to \mu^2/m^2)\Big\}
\end{align}
where $s=p^2/m^2$.

In Fig.~\ref{fig2}, we compare these expressions for the SU(2) gauge
group with the lattice simulations of~\cite{Cucchieri}. The best
choice of parameter is $g=7.5$ and $m=0.68$~GeV when normalization
prescriptions are imposed at $\mu=1$~GeV. One observes that both gluon
and ghost propagators can be fitted with the same choice of parameters
in a very satisfactory way. Note that the normalization conditions of
the lattice simulations are not compatible with~(\ref{rencond}) so
that we have to introduce a global multiplicative renormalization
factor when comparing the curves.

We have also compared our results with the data of two different
lattice studies~\cite{Bogolubsky09,Dudal10} for the SU(3) group. The
two data sets have different overall momentum scale and we have
rescaled the momenta of the data of~\cite{Bogolubsky09} for
superimposing them with those of~\cite{Dudal10}. We represent in
Fig.~\ref{fig3} the dressing function of the gluon instead of the
propagator in order to make visible the ultraviolet regime. The best
choice of parameters is $g=4.9$ and $m=0.54$~GeV (again with $\mu=1$
GeV) and it leads to a very satisfying agreement for momenta $p
\lesssim 2$~GeV. It is important to stress that
expressions~(\ref{4dprops}) are 1-loop results obtained from a fixed
coupling constant calculation in a fixed renormalization point. It is
well-known that in order to analyze the regime $p \gg m$, one must
take into account renormalization group effects and in particular the
running of the coupling. The corresponding procedure is standard and
once it is implemented (see \cite{Tissier10} for details), the
agreement is essentially within error bars for $p>m$ as is also shown
in Fig.~\ref{fig3}. A very good agreement is also obtained for the
ghost dressing function~\cite{Tissier10}. In any case, it is obvious
that when $p\gg m$, the model~(\ref{eq_lagrang}) reproduces correctly
the high momentum regime once renormalization group effects are taken
into account.

An interesting feature of the 1-loop gluon propagator is that it is
non-monotonous in the IR. In fact, the inverse propagator behaves at
small momenta as $m^2+N
g^2p^2/(192\pi^2)\log(p^2/m^2)+\mathcal{O}(p^2)$.  This prediction of
our calculation is very small for $d=4$ and it is not visible in
Fig.~\ref{fig2} but appears clearly in $d=3$, see below.

\begin{figure}
\epsfxsize=7.4cm
\epsfbox{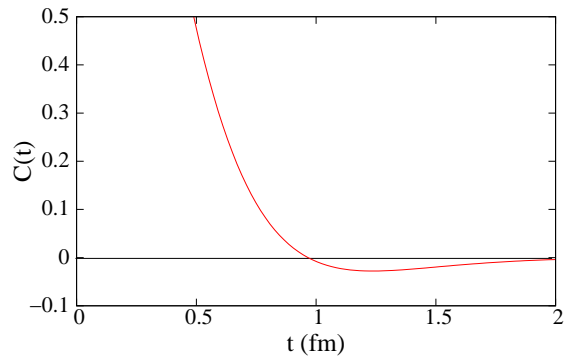}
\caption{\label{fig4}Four-dimensional real space propagator $C(t)$ for
  SU(3) gauge group. The curves grows when $t$ tends to zero,
  saturating at the value $C(0)\simeq 2.1$.}
\end{figure}

An important property of the propagators measured on the lattice is
the violation of positivity.  One way to extract it is to calculate
the quantity:
\begin{equation}
 C(t)=\int_{-\infty}^{\infty} \frac{dp}{2\pi} \mathrm{e}^{i pt} G(p).
\end{equation}
It can be shown (see for instance~\cite{Cucchieri_04}) that the
positivity of the spectral function implies the positivity of
$C(t)$. In Fig.~\ref{fig4}, we plot a numerical Fourier transform of
the SU(3) gluon propagator for the best parameters described above:
this shows a clear violation of positivity. We observe that the curve
of $C(t)$ is very similar to the one of~\cite{Bowman07}: it is
strongly positive for $t\lesssim $1 fm and slightly negative beyond.
\begin{figure}
\epsfxsize=7.4cm
\epsfbox{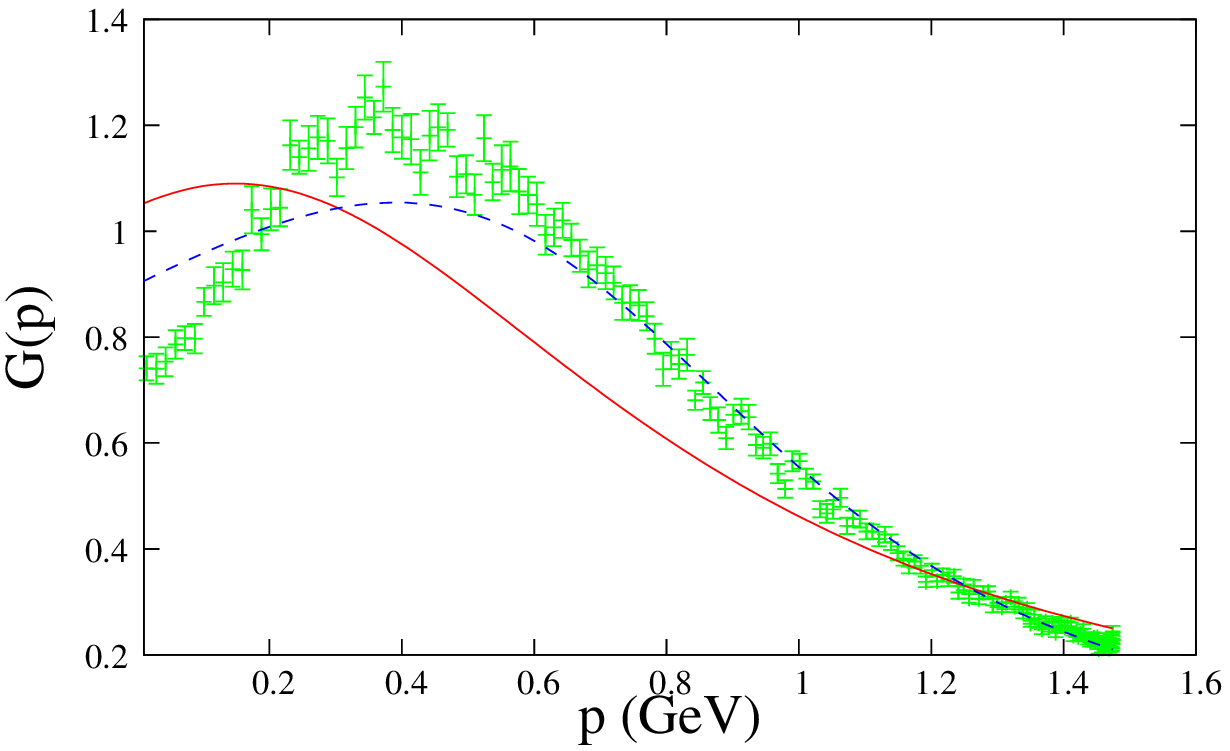}
\epsfxsize=7.4cm
\epsfbox{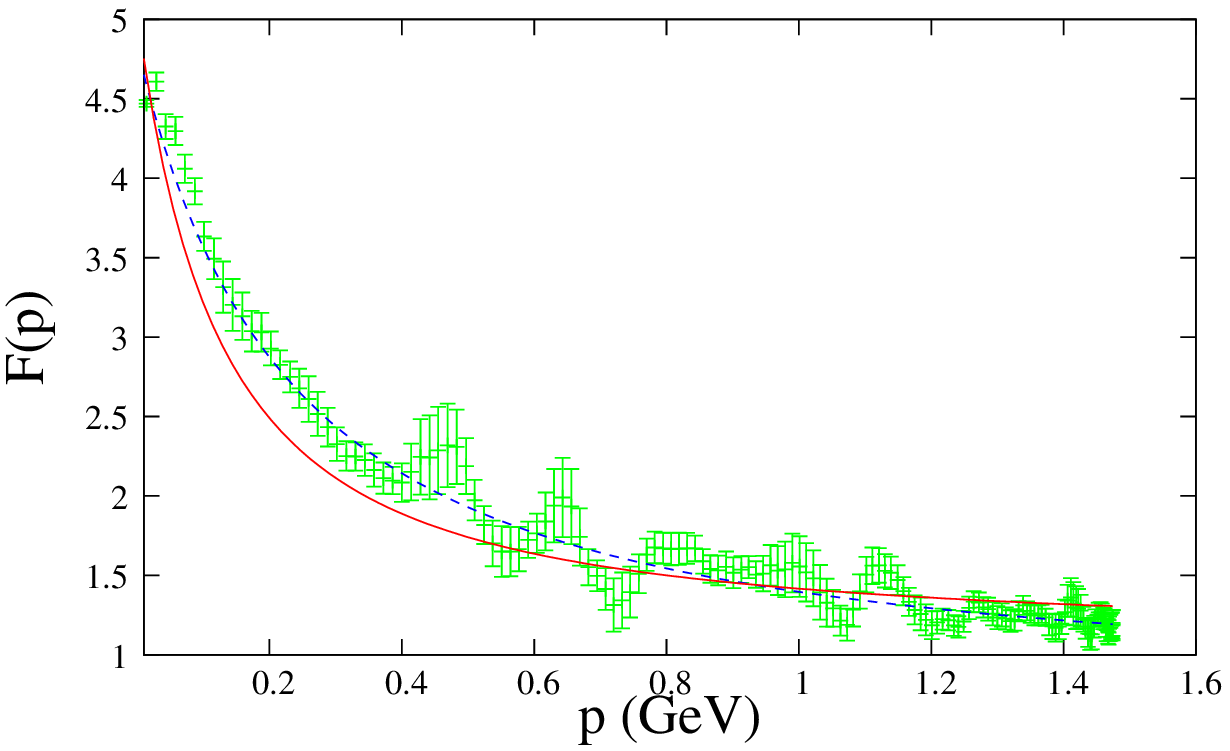}
\caption{\label{fig5}Three-dimensional functions for SU(2) gauge
  group: Comparison of present results (plain red curve for $\mu=$
  1~GeV and dashed blue curve for $\mu=$ 11~GeV) with lattice data
  of~\cite{Cucchieri} (green bars).  Top figure: gluon
  propagator. Bottom figure: ghost dressing function.}
\end{figure}
 
Let us now consider the three-dimensional case, where the 1-loop
calculation can be done explicitly again. The details of the
calculations and the final expressions will be presented in a future
publication~\cite{Tissier10}. We only mention here that the model
(\ref{eq_lagrang}) is able to account for the main features of gluon
and ghost propagators found in lattice simulations. In Fig.~\ref{fig5}
the results of the present model with the best fit parameters
$g=3.7$~$\sqrt{\text{GeV}}$ and $m=0.89$~GeV for $\mu=1$~GeV are
compared with $d=3$ simulations performed with the gauge group
SU(2)~\cite{Cucchieri}. We observe that the best fit for gluon and
ghost propagators are not as good as in $d=4$. This is probably
related to the fact that higher loop corrections are not very
small. It is worth mentioning that the results improve if one imposes
the normalization conditions at a larger momentum scale (for
$\mu=11$~GeV the best parameters are $g=1.6$~$\sqrt{\text{GeV}}$ and
$m=0.35$~GeV). Such a large scheme dependence indicates that higher
loop corrections give significant contributions. In any case, our
calculation reproduces the finite IR gluon propagator and ghost
dressing function. It also reproduces the non-monotonic behaviour of
the gluon propagator in the IR. An expansion of the inverse propagator
at low momentum leads to $m^2-Ng^2p/64+\mathcal{O}(p^2)$.

Within the model, the difference between $d=2$ and $d>2$ that is
observed in the lattice (see above) also appears natural.  In $d=2$,
we find that the gluon mass $m$ and ghost dressing function at zero
momentum $F(0)$ develop logarithmic IR divergences. Such divergences
exclude the possibility of controlling the one-loop calculation as was
used above for $d>2$. A proper treatment of the $d=2$ case requires a
renormalization group approach adapted to the IR
regime~\cite{Tissier10} and goes beyond the scope of the present
letter.

The model presented here reproduces surprisingly well the $d=4$
propagators of pure gluodynamics for SU(2) and SU(3) and describes in
a simple way the main characteristics of those propagators in
$d=3$. The specificities of the $d=2$ case result from the IR
divergences that appear in this dimension. Given the technical
simplicity of this approach, this work opens the door to many
subsequent applications in Strong Interactions physics (for example,
the inclusion of quarks, a study of the dependence on the gauge
fixing, three and four point correlation functions, quark-anti-quark
static potential).

Considering the surprisingly good agreement between the 1-loop
calculation and the simulations, it is tempting to think that the
model is not just a good phenomenological description and one should
try to justify the use of this action from first principles.  The
issue of unitarity should be also explored.

{\it Acknowledgments.} We thank B. Delamotte, A. Cucchieri,
O. Oliveira, M. Mueller-Preussker, A. Sternbeck, and G.Tarjus for
useful discussions. M.T. thank the IFFI for its
hospitality. N.W. acknowledge the support of the PEDECIBA program.


\vspace{-.3cm}
\begin{thebibliography}{10}

\bibitem{Montvay94}
  I.~Montvay and G.~Munster,
  ``Quantum fields on a lattice,''
{\it  Cambridge, UK: Univ. Pr. (1994) 
}

\bibitem{Cucchieri}
  A.~Cucchieri and T.~Mendes,
  Phys.\ Rev.\ Lett.\  {\bf 100} (2008) 241601 and also arXiv:1001.2584 [hep-lat].

\bibitem{Bogolubsky09}
  I.~L.~Bogolubsky {\it et al.},
  Phys.\ Lett.\  B {\bf 676}, 69 (2009).


\bibitem{Dudal10}
  D.~Dudal, O.~Oliveira and N.~Vandersickel,
  arXiv:1002.2374 [hep-lat].


\bibitem{Cucchieri_08}   
A.~Cucchieri and T.~Mendes,   
Phys.\ Rev.\  D {\bf 78} (2008) 094503   [arXiv:0804.2371 [hep-lat]].   

\bibitem{Maas_07}   A.~Maas,   
Phys.\ Rev.\  D {\bf 75} (2007) 116004   

\bibitem{Bowman07}
  P.~O.~Bowman {\it et al.},
  Phys.\ Rev.\  D {\bf 76}, 094505 (2007).

\bibitem{Cucchieri_04}
  A.~Cucchieri, T.~Mendes and A.~R.~Taurines,
  Phys.\ Rev.\  D {\bf 71} (2005) 051902.

\bibitem{Gribov77}
  V.~N.~Gribov,
  Nucl.\ Phys.\  B {\bf 139} (1978) 1.

\bibitem{Zwanziger89}
  D.~Zwanziger,
  Nucl.\ Phys.\  B {\bf 323}, 513 (1989).

\bibitem{Zwanziger92}
  D.~Zwanziger,
  Nucl.\ Phys.\  B {\bf 399}, 477 (1993).


\bibitem{Dudal08}
  D.~Dudal, J.~A.~Gracey, S.~P.~Sorella, N.~Vandersickel and H.~Verschelde,
  Phys.\ Rev.\  D {\bf 78} (2008) 065047.

\bibitem{Alkofer00}
  R.~Alkofer and L.~von Smekal,
  Phys.Rept.{\bf 353} (2001) 281.

\bibitem{Fischer04}
  C.~S.~Fischer and H.~Gies,
  JHEP {\bf 0410} (2004) 048.

\bibitem{Boucaud06}
  Ph.~Boucaud {\it et al.},
  JHEP {\bf 0606} (2006) 001.

\bibitem{Curci76}
  G.~Curci and R.~Ferrari,
  Nuovo Cim.\ A {\bf 32}, 151 (1976).

\bibitem{Natale09}
  A.~A.~Natale,
  arXiv:0910.5689 [hep-ph].

\bibitem{Cornwall79}
  J.~M.~Cornwall,
  Nucl.\ Phys.\  B {\bf 157}, 392 (1979).



\bibitem{BRST}
  C.~Becchi, A.~Rouet and R.~Stora,
  Commun.\ Math.\ Phys.\  {\bf 42}, 127 (1975) and
Annals Phys.\  {\bf 98} (1976) 287,
I.~V.~Tyutin, LEBEDEV-75-39.

\bibitem{Delduc89}
  F.~Delduc and S.~P.~Sorella,
  Phys.Lett.B{\bf 231}, 408 (1989).

\bibitem{Tissier08}
  M.~Tissier and N.~Wschebor,
  Phys.\ Rev.\  D {\bf 79}, 065008 (2009).

\bibitem{Dudal02}
  D.~Dudal, H.~Verschelde and S.~P.~Sorella,
  Phys.\ Lett.\  B {\bf 555} (2003) 126.

\bibitem{Taylor71}
J.~C.~Taylor,
Nucl.\ Phys.\ B {\bf 33} (1971) 436.



\bibitem{Tissier10}
  M.~Tissier and N.~Wschebor, {\it in preparation}.


\end{thebibliography}
\end{document}